\documentclass[fleqn,usenatbib]{mnras}

\usepackage{newtxtext,newtxmath}

\usepackage[T1]{fontenc}
\usepackage{ae,aecompl}

\usepackage{graphicx}	
\usepackage{amsmath}	
\usepackage{amssymb}	
\usepackage{xspace}
\usepackage{multirow}

\newcommand{\mic}{$\mu$m\xspace}

\newcommand{\lsd}{\hbox{$\lambda/D$}\xspace}

\title[Inner structures around HD169142]{Investigation of the inner structures around HD169142 with VLT/SPHERE}

\author[R. Ligi et al.]{
R. Ligi,$^{1}$\thanks{E-mail: roxanne.ligi@lam.fr}
A. Vigan,$^{1}$
R. Gratton,$^{2}$
J. de Boer,$^{3}$ 
M. Benisty,$^{4}$
A. Boccaletti,$^{5}$ 
\newauthor
S. P. Quanz,$^{6}$ 
M. Meyer,$^{6,7}$ 
C. Ginski,$^{3}$ 
E. Sissa,$^{2}$
C. Gry,$^{1}$
T. Henning,$^{8}$ 
J.-L. Beuzit,$^{4}$
\newauthor
B. Biller,$^{9}$ 
M. Bonnefoy,$^{4}$
G. Chauvin,$^{4}$
A. C. Cheetham,$^{10,11}$
M. Cudel,$^{4}$
P. Delorme,$^{4}$
\newauthor
S. Desidera,$^{2}$
M. Feldt,$^{8}$ 
R. Galicher,$^{5}$ 
J. Girard,$^{12,4}$
M. Janson,$^{13}$
M. Kasper,$^{14,4}$
\newauthor
T. Kopytova,$^{8,15,16}$ 
A.-M. Lagrange,$^{4}$
M. Langlois,$^{17,1}$
H. Lecoroller,$^{1}$
A.-L. Maire,$^{8}$ 
\newauthor
F. M\'enard,$^{4}$
D. Mesa,$^{2}$
S. Peretti,$^{10}$
C. Perrot,$^{5}$ 
P. Pinilla,$^{18}$
A. Pohl,$^{8,19}$ 
D. Rouan,$^{5}$ 
\newauthor
T. Stolker,$^{20}$
M. Samland,$^{8}$ 
Z. Wahhaj,$^{12,1}$ 
F. Wildi,$^{10}$
A. Zurlo,$^{21,22,1}$ 
T. Buey,$^{5}$ 
\newauthor
D. Fantinel,$^{2}$
T. Fusco,$^{23}$
M. Jaquet,$^{1}$
T. Moulin,$^{4}$
J. Ramos,$^{8}$ 
M. Suarez,$^{14}$
L. Weber$^{10}$\\
$^{1}$Aix Marseille Univ, CNRS, LAM, Laboratoire d'Astrophysique de Marseille, Marseille, France \\
$^{2}$INAF -- Osservatorio Astronomico di Padova, Vicolo dell'Osservatorio 5, I-35122, Padova, Italy\\
$^{3}$Leiden Observatory, Leiden University, PO Box 9513, 2300 RA Leiden, The Netherlands \\
$^{4}$Univ. Grenoble Alpes, CNRS, IPAG, 38000 Grenoble, France \\
$^{5}$LESIA, Observatoire de Paris, PSL Research University, CNRS, Sorbonne Universit\'es, UPMC Univ. Paris 06, Univ. Paris Diderot, \\ Sorbonne Paris Cit\'e  \\
$^{6}$Institute for Astronomy, ETH Zurich, Wolfgang-Pauli-Strasse 27, 8093 Zurich, Switzerland\\
$^{7}$Department of Astronomy, University of Michigan, 1085 S. University Ave, Ann Arbor, MI 48109-1107, USA \\
$^{8}$Max Planck Institute for Astronomy, K\"onigstuhl 17, D-69117 Heidelberg, Germany \\
$^{9}$Institute for Astronomy, University of Edinburgh, Blackford Hill, Edinburgh EH9 3HJ, UK; bb@roe.ac.uk \\
$^{10}$Observatoire de Gen\`eve, Universit\'e de Gen\`eve, 51 chemin des Maillettes, 1290, Versoix, Switzerland\\
$^{11}$Sydney Institute for Astronomy, School of Physics, University of Sydney, NSW 2006, Australia\\
$^{12}$European Southern Observatory (ESO), Alonso de C\`ordova 3107, Vitacura, 19001 Casilla, Santiago, Chile\\
$^{13}$Department of Astronomy, Stockholm University, SE-106 91 Stockholm, Sweden\\
$^{14}$European Southern Observatory, Karl-Schwarzschild-Str. 2, D85748 Garching, Germany \\
$^{15}$School of Earth $\&$ Space Exploration, Arizona State University, Tempe AZ 85287, USA\\
$^{16}$Ural Federal University, Yekaterinburg 620002, Russia\\
$^{17}$CRAL, UMR 5574, CNRS, Universit\'e Lyon 1, 9 avenue Charles Andr\'e, 69561 Saint Genis Laval Cedex, France \\
$^{18}$Department of Astronomy/Steward Observatory, The University of Arizona, 933 North Cherry Avenue, Tucson, AZ 85721, USA \\
$^{19}$Heidelberg University, Institute of Theoretical Astrophysics, Albert-Ueberle-Str. 2, D-69120 Heidelberg, Germany \\
$^{20}$Anton Pannekoek Institute for Astronomy, University of Amsterdam, Science Park 904, 1098 XH Amsterdam, The Netherlands \\
$^{21}$N\'ucleo de Astronom\'ia, Facultad de Ingenier\'ia, Universidad Diego Portales, Av. Ejercito 441, Santiago, Chile \\
$^{22}$Millenium Nucleus "Protoplanetary Disk", Departamento de Astronomi\'a, Universidad de Chile, Casilla 36-D, Santiago, Chile \\
$^{23}$ONERA, 29 avenue de la Division Leclerc, F-92320 Ch\^atillon, France
}

\date{Accepted 2017 September 03. Received 2017 September 03; in original form 2017 March 02}

\pubyear{2017}

\begin{document}

\label{firstpage}
\pagerange{\pageref{firstpage}--\pageref{lastpage}}
\maketitle

\begin{abstract}
We present observations of the Herbig Ae star HD169142 with VLT/SPHERE instruments InfraRed Dual-band Imager and Spectrograph (IRDIS) ($K1K2$ and $H2H3$ bands) and the Integral Field Spectrograph (IFS) ($Y$, $J$ and $H$ bands). We detect several bright blobs at $\sim$180~mas separation from the star, and a faint arc-like structure in the IFS data. Our reference differential imaging (RDI) data analysis also finds a bright ring at the same separation. We show, using a simulation based on polarized light data, that these blobs are actually part of the ring at 180~mas.  These results demonstrate that the earlier detections of blobs in the $H$ and $K_S$ bands at these separations in Biller et al. as potential planet/substellar companions are actually tracing a bright ring with a Keplerian motion.
Moreover, we detect in the images an additional bright structure at $\sim$93~mas separation and position angle of 355$^{\circ}$, at a location very close to previous detections. It appears point-like in the $YJ$ and $K$ bands but is more extended in the $H$ band. We also marginally detect an inner ring in the RDI data at $\sim$100~mas. Follow-up observations are necessary to confirm the detection and the nature of this source and structure.
\end{abstract}




\section{Introduction} 
\label{sec:Introduction}

Young stellar objects are surrounded by circumstellar material, making them ideal targets to study planetary formation. Transitional disks are particularly interesting as they may constitute the intermediate step between gas-rich protoplanetary disks where planets are supposed to form, and dusty debris disks. \\ 
Direct observations of companions and disk structures are necessary to bring constraints on planetary formation. A few targets have already been identified as interesting cases to study this phenomenon, such as HD\,100546 \citep{Quanz2015,Quanz2013,Brittain2013,Currie2015}, HD\,142527 \citep{Biller2012}, or LkCa~15 \citep{Kraus2012, Sallum2015}. These examples show that determining the origin of disk structures is a difficult task, and overall, the risk to confuse them with forming planets is quite high \citep[see e.g.][]{Follette2017}.\\
HD\,169142 is a well studied Herbig Ae \citep{Meeus2010} star at 117~pc \citep[][see Tab.~\ref{tab:param}]{GaiaCat2016,Michalik2015} hosting a nearly face-on disk often categorized as pre-transitional since it shows dust emissions both at close and large separations from the star separated by several gaps \citep{Wagner2015,Osorio2014}.
The disk has first been spatially resolved by \citet[][$H$ band]{Kuhn2001} with polarimetry and studied by \citet[][2-45~$\mu m$]{Meeus2001} with spectroscopy, and later confirmed by \citet[][$JHK$ bands]{Hales2006}. \citet{Quanz2013} reported polarimetric observations with NaCo in $H$ band, revealing a bright irregular ring at 170~milliarseconds (mas), that is 20~AU, and an annular gap from 270 to 480~mas (32-56~AU), the surface brightness smoothly decreasing after 550~mas (66~AU). \citet{Monnier2017} confirm a double ring structure using the Gemini Planet Imager (GPI) in $H$ band, showing a surface brightness enhancement at 180~mas (21 AU) and one at 510~mas. Interestingly, ALMA (Atacama Large Millimetre/Sub-millimeter Array) observations revealed two rings at 170-300~mas and 479-709~mas, and an empty cavity (R<171~mas) at the center of the dust disk but filled with gas \citep{Fedele2017}. This might indicate the presence of multiple planets carving out the gaps and cavities \citep{Zhu2012, Dong2015}, or alternatively that magneto-rotational instability (MRI) effects combined with magnetohydrodynamical (MHD) winds \citep{Pinilla2016} shape the disk density structure.

Additional hints of planet formation have been identified around HD\,169142. \citet{Biller2014} and \citet{Reggiani2014} performed observations of HD\,169142 with NaCo in the $L'$ band. \citet{Biller2014} detected a faint marginally resolved point-like feature in the data from July 2013, located at a position angle (PA) of 0$\pm14^\circ$ and a separation of $\rho$=110$\pm$30~mas (13$\pm$3.5 AU), with $\Delta mag$=6.4$\pm$0.2. If this emission was photospheric, it would correspond to a 60-80~$M_{\rm Jup}$ brown dwarf companion. However, this companion was not confirmed by shorter wavelength follow-up observations performed with the adaptive optics system at the Magellan Clay Telescope (MagAO/MCT) in $H$, $K_S$ and $z_p$ bands (where it should have been easily detected if it was a 60-80 M$_{\rm Jup}$ companion), nor at 3.9$\mu m$, though at lower sensitivity. This suggests that the object found in 2013 might be a part of the disk possibly showing planetary formation with an unknown heating source. \cite{Reggiani2014} also detected a point-souce in NaCo data from June 2013. This emission source was at $\rho=$156$\pm$32~mas (18$\pm$3.8 AU) and PA=7.4$\pm$1.3$^\circ$. It has a $\Delta mag$=6.5$\pm$0.5, and an apparent magnitude of 12.2$\pm$0.5~mag. They suggest that this could come from the photosphere of a 28-32~$M_{\rm Jup}$ companion, or from an accreting lower-mass forming planet in the gap. Additional observations were carried out with the GPI instrument \citep{Macintosh2014} in the $J$ band in April 2014, where this hypothetical companion was not retrieved, again suggesting that it was not a 28-32~M$_{\rm Jup}$ object, as this should have been relatively easily detected in $J$ band.
Finally, \citet{Osorio2014} additionally find with the Expanded Very Large Array (EVLA) 7~mm observations a knot of emission at 350~mas (41~AU at 117~pc), that could correspond to an object of 0.6~M$_{\rm Jup}$.

\begin{figure}
\centering
\includegraphics[scale=0.6]{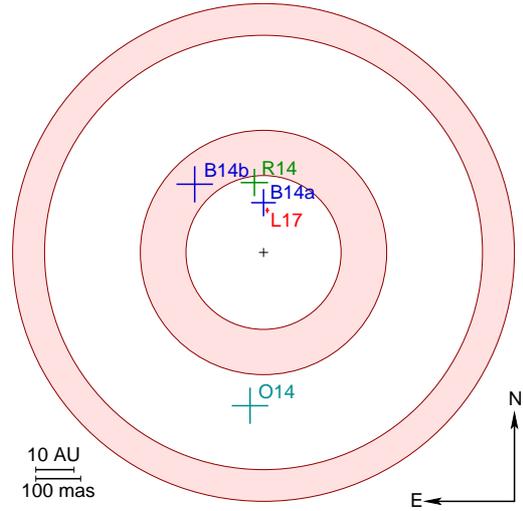}
\caption{Diagram of the HD169142 system. The red parts represent the two rings and the white parts are the gaps. The crosses represent the positions of the point-like structures discovered by \citet[][B14b for the structure discovered with MagAO and B14a for the one detected with NaCo]{Biller2014}, \citet[][R14]{Reggiani2014} and \citet[][O14]{Osorio2014}. We also show the structure around 100~mas detected with SPHERE in this work (L17, see Sec.~\ref{sec:Structure}). The dimension of the crosses represent the error bars (on scale), except for B14b and O14 where the error bars are given arbitrarily. The inclination of the disk is not shown in this diagram.}
\label{fig:Schema}
\end{figure}

The follow-up MagAO/MCT observations performed by \citet{Biller2014} led to another, low signal-to-noise ratio (S/N) detection at $\rho=180$~mas (21 AU) and PA=33$^\circ$. If a real companion, this structure would correspond to a 8-15~$M_{\rm Jup}$ substellar companion, but it was not found in the initial 2013 NaCo data in $L'$ band. 
All these results demonstrate the complexity of this system which is even more critical given the lack of consistency (possibly because of observational limitations) between the results. Figure~\ref{fig:Schema} summarises the different point-like structures identified around HD169142 so far. \\

In this paper, we investigate the innermost structures (<300~mas) previously detected around HD\,169142 to confirm the presence of the candidate companions detected by \cite{Biller2014} and \cite{Reggiani2014} and investigate their nature. We present new near infrared (NIR) observations of HD\,169142 obtained with SPHERE/VLT \citep[Spectro-Polarimetric High-contrast Exoplanet REsearch,][]{Beuzit2008}, as part of the Guaranteed Time Observations (GTO) dedicated to exoplanet search (SpHere INfrared survey for Exoplanets or SHINE, Chauvin et al. in prep.). SPHERE has primarily been designed to image and characterize exoplanets, but it is also a powerful instrument for probing the dusty surface of protoplanetary disks.
The observations are described in Sec.~\ref{sec:Obs} and the data analysis in Sec.~\ref{sec:datanalysis}. We report in Sec.~\ref{sec:Blobs} the detection of bright blobs at 180~mas that are actually part of the inner ring, and we show in Sec.~\ref{sec:Structure} the marginal detection of a bright structure located at similar position to the object found by \cite{Biller2014} and \cite{Reggiani2014}. We conclude in Sec.~\ref{sec:conclusion}.

\begin{table}
	\centering
	\caption{Parameters of HD169142.}
	\label{tab:param}
	\begin{tabular}{l c c}	
	    \hline
    	Parameter & Value 	& Ref. \\
	    \hline
	    RA (J2000)&	18$^{\rm h}$24$^{\rm min}$29$^{\rm s}$.787		&	(1)	\\ 
	    DEC (J2000) &	 -29$^\circ$46'49''.22	&	(1)	\\ 
	    Parallax [mas]	&	8.526$\pm$0.288	&	(2) 	\\ 
	    Distance [pc]	&	117.288$^{+3.832}_{-4.101}$	&	(3)	\\ 
    	J	[mag]			&	7.31$\pm$0.02	&	(1)	\\ 
    	H	[mag]			&	6.91$\pm$0.04	&	(1)	\\ 
    	K$_S$	[mag]			&	6.41$\pm$0.02	&	(1)\\ 
    	L'	[mag]			&	5.66$\pm$0.03	&	(4)		\\ 
    	G	[mag]			& 8.060				& (2)  \\ 
    	Age [Myr]			&	 1-5;6$^{+6}_{-3}$;12	&	(5)\,; (6)\,;(7)	\\
    	M	[M$_\odot$]			&	1.65	&	(7) \\
    	R	[R$_\odot$]				&	1.59;$\sim$1.6	&	(7)\,;(8)	\\
    	L	[L$_\odot$]				&	8.55 & (7) \\
    	T$_{\rm eff}$ [K] &	7500$\pm$200;6464;7500-7800 & (5)\,;(7)\,;(8) \\
    	Spec. Ty. &  A9III/IVe ;A7V  & 	(5)\,;(7)		\\
    	Fe/H & -0.5$\pm$0.1;-0.5-0.25 & (5)\,;(8) \\
    	$\log(g)$  & 3.7$\pm$0.1;4-4.1  & (5)\,;(8) \\
    	\hline
	\end{tabular}
\textbf{References}. (1) 2MASS catalog \citep{CutriCat2003}\,; (2) Gaia Catalog \citep{GaiaCat2016,Michalik2015}\,; (3) adapted from Gaia catalog \citep{GaiaCat2016}\,; (4) \cite{Malfait1998}\,; (5) \cite{Guimaraes2006}\,; (6) \cite{Grady2007}\,; (7) \cite{Blondel2006} \,; (8) \cite{Meeus2010}.
\end{table}

\section{Observations and data reduction}
\label{sec:Obs}

\begin{table*}
	\centering
	\caption{Observing log of SPHERE SHINE data for HD169142. Platescales values for IRDIS are given for the $K1K2$ filters or $H2H3$ filters.}
	\label{tab:obslog}
	\begin{tabular}{lcccccccc} 
		\hline
		UT Date		 & MJD 	&	Corona-	& Instr.			&  DIT$\times$NDITs	&	Exposure time &	Field rotation	&	Mean seeing & Platescale \\
							 &  [day]	&	graph	&	$\&$  	Band			&					&			  [s]					&  [deg]			& [''] &	 [mas]\\
		\hline						 
		\multirow{2}{*}{2015 June 7} &	\multirow{2}{*}{57180.17}	& \multirow{2}{*}{Y} &	IFS YJ 	&		64.0$\times$86		&	91.7	&\multirow{2}{*}{45.82}	& \multirow{2}{*}{1.57}	 &7.46$\pm$0.02\\
														&  												&								&	IRDIS H2H3	&		64.0$\times$4 or 2	&			102.4	&							&				&	 12.255/12.250$\pm$0.009\\
			
		\hline
				\multirow{2}{*}{2015 June 28} &	\multirow{2}{*}{57201.12}	& \multirow{2}{*}{Y} &	IFS YJH	&		64.0$\times$65		&	69.3	& \multirow{2}{*}{36.42}	& \multirow{2}{*}{1.00}	 &7.46$\pm$0.02\\
														&  												&								&	IRDIS	 K1K2	&		64.0$\times$5	&		85.33			&						&		&12.267/12.263$\pm$0.009\\
		\hline
				\multirow{2}{*}{2016 April 21} &	\multirow{2}{*}{57499.34}	& \multirow{2}{*}{Y} &	IFS YJ 	&	64.0$\times$77	&	82.1	&	\multirow{2}{*}{145.0}	& \multirow{2}{*}{1.88} & 7.46$\pm$0.02\\
														&  												&								&	 IRDIS H2H3	&		64.0$\times$17	&	90.67	&							&							&	 12.255/12.250$\pm$0.009\\
		\hline
			\multirow{2}{*}{2016 June 27} &	\multirow{2}{*}{57566.15}	& \multirow{2}{*}{N} &	IFS 	YJH		&		2.0$\times$1610	&64.8 	&	\multirow{2}{*}{149.9} & \multirow{2}{*}{0.67}	& 7.46$\pm$0.02\\
														& 													 &									& IRDIS K1K2	&		0.84$\times$38	&			64.22 	&								&  				&	12.249/12.245$\pm$0.009\\
		\hline
		\multirow{2}{*}{2017 April 30} &	\multirow{2}{*}{57873.30}	& \multirow{2}{*}{N} &		IFS YJH		&		2.0$\times$1152	&61.2	&	\multirow{2}{*}{98.82}	&		\multirow{2}{*}{0.62}		&	7.46$\pm$0.02\\
														&													  &								& 		IRDIS 	K1K2	&	0.84$\times$561	&		78.10							&			 			&			&	12.249/12.245$\pm$0.009\\
		\hline
	\end{tabular}
	 Notes: The DIT values refer to the intergration time, and the NDIT values to the number of integration per datacube.
\end{table*}

Observations of HD\,169142 were performed from 2015 to 2017 (see Tab.~\ref{tab:obslog}). The data were obtained in the \texttt{IRDIFS} or in the \texttt{IRDIFS$\_$EXT} modes, using simultaneously the DBI \citep[Dual-Band Imaging,][]{Vigan2010} mode. For the \texttt{IRDIFS} mode, the Integral Field Spectrograph \citep[IFS,][]{Claudi2008} was operating in the wavelength range between 0.95~\mic and 1.35~\mic ($YJ$) at a spectral resolution of R=50, and the InfraRed Dual-band Imager and Spectrograph \citep[IRDIS,][]{Dohlen2008} in the $H$ band with the H23 filter pair ($\lambda_{\mathrm{H2}} = 0.055$~\mic, $\lambda_{\mathrm{H3}} = 1.667$~\mic). For the \texttt{IRDIFS$\_$EXT} mode, the IFS was used between 0.95~\mic and 1.65~\mic ($YJH$) (R=30) and IRDIS in the $K$ band with the K12 filter pair ($\lambda_{\mathrm{K1}} = 2.110$~\mic, $\lambda_{\mathrm{K2}} = 2.251$~\mic).
Due to very small angular separation of previously reported detections, the two most recent observations were done without coronagraph, but the core of the stellar point-spread function (PSF) was saturated over a radius of $\sim$1~\lsd and observations were performed in pupil-stabilized mode to enable angular differential imaging \citep[ADI,][]{Marois2006}. 

To check the consistency of the results, different pipelines were used to reduce and analyse the IFS and IRDIS data. We used the LAM-ADI pipeline \citep{Vigan2015,Vigan2016} pipeline and the SPHERE Data Reduction and Handling (DRH) automated pipeline \citep{Pavlov2008} for IRDIFS data, and the pipeline described in \citet[][ASDI-PCA algotrithm]{Mesa2015} for the IFS data. Even though the observing conditions were good, there were some temporal variations so we performed a frame selection on the data sets.   
We used the sortframe routine developed by the SPHERE Data Center (DC) to select the good frames when using the DC and \cite{Mesa2015} pipelines. The minimum fraction of selected frames is about 80$\%$ and the maximum one is near 100$\%$, depending on weather conditions. For the LAM-ADI pipeline, we calculated the moving average of the flux in an annulus centered on the star. We then excluded the frames presenting a flux above or below 1.5$\sigma$ of the mean flux. This method follows a Gaussian behavior and corresponds to $\sim$14$\%$ of the frames removed using the 1.5$\sigma$ criterion. This allows to keep enough frame to have a correct S/N while removing the very bad frames that could induce artifacts in the images (this case only applies to the images shown in Fig~\ref{fig:simu}). 
Finally, the SHINE data were astrometrically calibrated following the analysis in \cite{Maire2016}.
To improve the S/N, and to show the different structures appearing in the images, the selected IFS and IRDIS data were collapsed to broad-band images equivalent to $K$, $H$ and $YJ$ bands.

\section{Data analysis}
\label{sec:datanalysis}

\subsection{PCA reduction}
\label{sec:PCAreduction}

\begin{figure*}
	\includegraphics[width=0.9\textwidth]{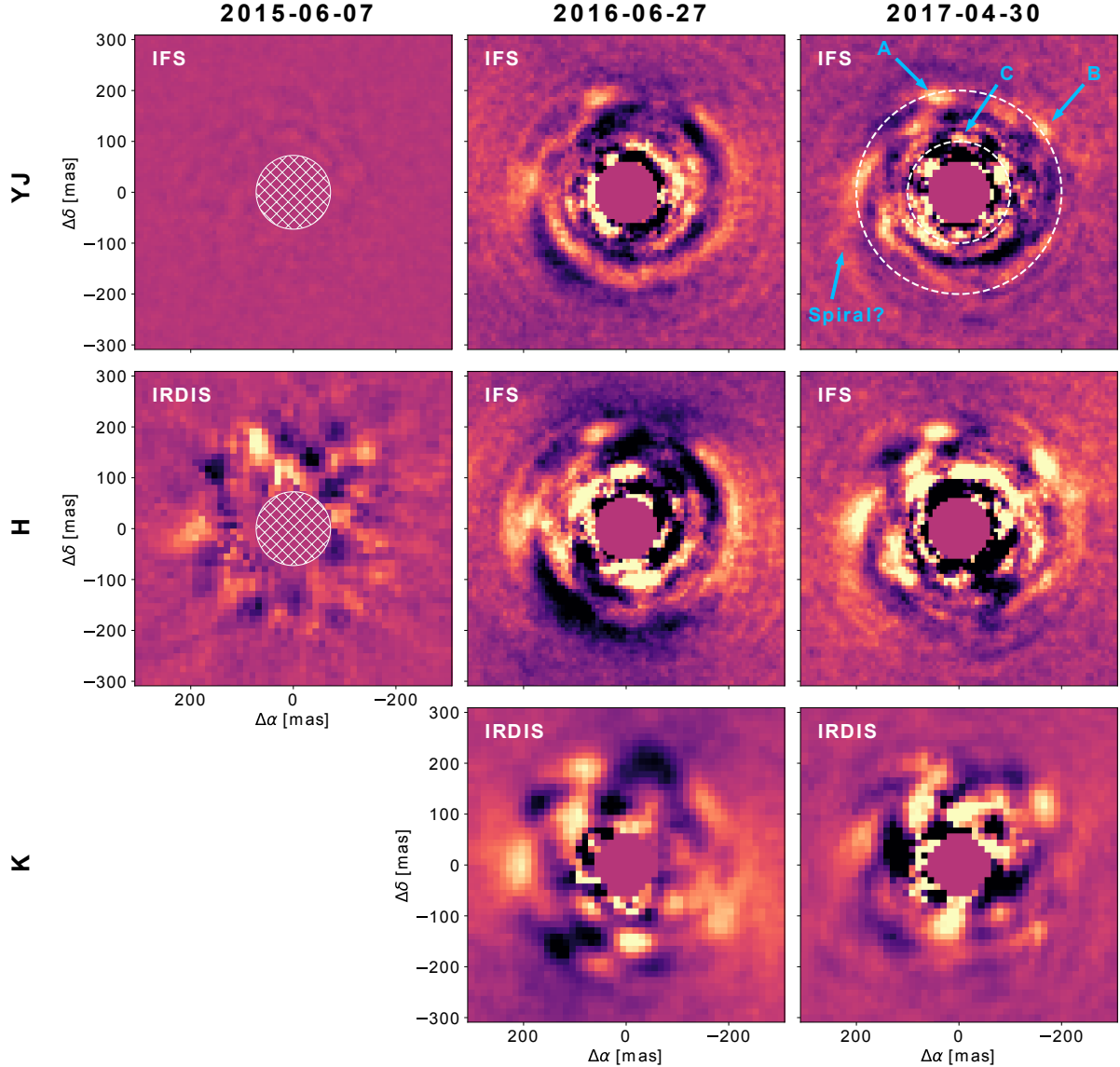}
	\caption{Result of the PCA analysis on the IFS \citep[][pipeline]{Mesa2015} with 50 subtracted modes, and IRDIS \citep[][pipeline]{Vigan2015} with 6, 50 and 20 subtracted modes for 2015, 2016 and 2017 data, respectively. The square-pattern central circle shows the position of the coronagrah in the 2015 data set; the star is at the center. The bright structures are indicated with blue arrows. Letter A indicates the structure at PA = 20$^{\circ}$, and letter B indicates the one at 310$^{\circ}$, both being at separation $\sim$180~mas. Letter C shows the structure at $\sim$100~mas and PA = 355$^{\circ}$. The two white dashed circles have a radius of 100~mas and 180~mas, respectively. North is up and east is left.}
	\label{fig:detectionIRDIFS}
\end{figure*}

The data were first analysed using principal component analysis (PCA) based on the formalism described in \citet{Soummer2012}. The modes were calculated over the full sequence at separations up to 500~mas. A variable number of modes were subtracted, up to $\sim10\%$ of the total number of modes for IRDIS ($\sim$50), and up to 50 modes for the IFS, before rotating the images to a common orientation and combining them with an average. The resulting images obtained in $YJ$, $K$ and $H$ band for periods 2015 June 7 (best quality image for 2015), 2016 June 27 and 2017 April 30 (best quality images for the 2016 and 2017 periods) are presented in Fig.~\ref{fig:detectionIRDIFS}.\\
Figure~\ref{fig:detectionIRDIFS} shows extended and point-like surface brightness enhancements depending on the band, and a faint arc-like feature in IFS data appearing on the East part of the disk, in particular in $YJ$ band (refered to as spiral). The bright structures are located at separations of $\sim$180-200~mas, especially at PA = 20$^{\circ}$ (structure A), 90$^{\circ}$, and 310$^{\circ}$ (structure B). Other structures are detected at $\sim150$~mas (PA= 320$^{\circ}$) and at $\sim100$~mas (PA = 355$^{\circ}$; structure C) from the central star. In IRDIS data, the bright structures are still visible but appear fainter. The structures appear point-like in the $YJ$ and $K$ band and more extended in the $H$ band. They are persistent whatever the number of subtracted modes in IFS and IRDIS data.
In Appendix~\ref{sec:AppendixA}, we show the S/N maps of epochs 2017 April 30 and 2016 June 27. The maps show bright and dark structures with positive and negative values of the S/N. The S/N is calculated as the normalized difference in intensity between a considered feature and two neighbouring areas at the same separation \citep[following the method described in][]{Zurlo2014}. Thus, bright peaks are significant features, just as much as dark peaks since they indicate darker features than the surrounding background. It is important to note that the calculation of the S/N can be modulated by the background which is inhomogeneous. In particular, the dark peak at $\sim40^{\circ}$ in the S/N map is not dark in Fig.~\ref{fig:cadi_simulation_profile}, but is surrounded by two bright structures. The positions of the structures showing a high signal (S/N$\sim$3) are consistent with the structures appearing in Fig.~\ref{fig:detectionIRDIFS}, in particular structure A and B. Structure C appears with a lower S/N. 

Several features have already been discovered at the separations of structures A, B and C (a bright ring and candidate companions, see Sec.~\ref{sec:Introduction}), but they appeared point-like. Since we detect bright spots at similar positions in our images, we try to investigate whether or not these detections are the same as the previous ones. In the next section, we analyse the structures found around $\sim$180-200~mas, and in Sec.\ref{sec:Structure}, we focus on the detection at $\sim$100~mas.

\subsection{RDI reduction}

We performed Reference Differential Imaging \citep[RDI, ][]{Soummer2014} which consists in subtracting the reference image of one or several stars to the target image. This technique allows subtracting the speckle pattern, while limiting the self-subtraction effects usually affecting ADI data, in particular in the case of extended structures like disks \citep[e. g.][]{Milli2012}. To select the reference images, we searched in the complete database of SPHERE GTO observations the reduced images that have the best correlation with our target images. This means that we calculated the correlation coefficient between the data sets taken in a similar band as the considered target observation, and the considered data set of our target. The best correlation coefficient (that is > 0.90 in general) designates the data set that is used as reference image. For images taken without coronagrah, the correlation coefficient is lower (around 0.50) than with coronagraph because there are fewer images taken without the coronagraph in the SPHERE database. Similarly, there are many more $Y-J$ images than $Y-H$ images, leading to lower coefficients for the latter mode.

\begin{figure*}
	\includegraphics[width=0.7\textwidth]{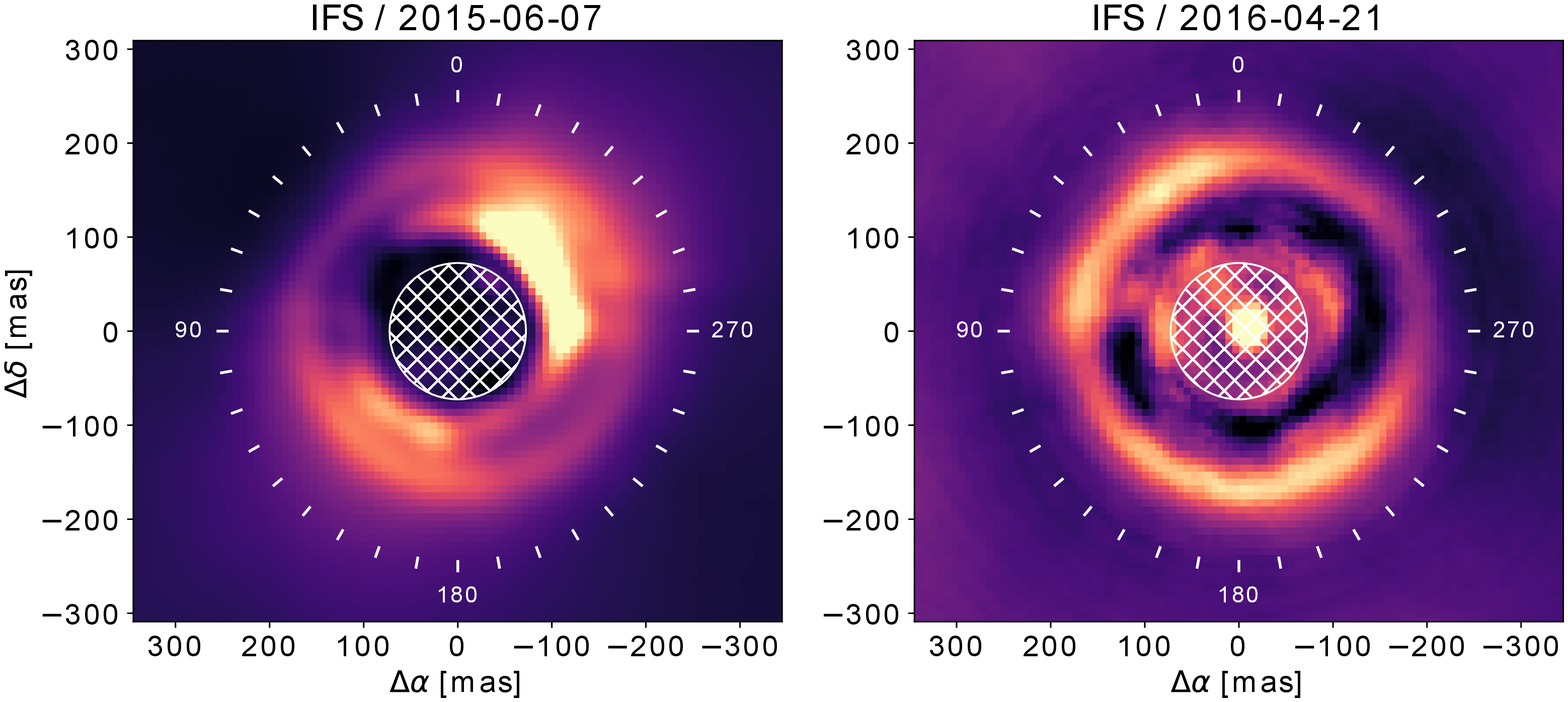}
	\caption{Result of the RDI analysis of the IFS data from 2015 June and 2016 April. We clearly see an inhomogenous bright ring at $\sim$180~mas, and possibly another inner ring, althought its position close to the star makes it less trustable. North is up and east is left.}
	\label{fig:RDI}
\end{figure*}

Figure~\ref{fig:RDI} shows the RDI IFS images of HD169142 from June 2015 and April 2016 with a subtracted image of the same night each time. We see a possible double ring structure, with one being located at $\sim$180 mas, and possibly another one at $\sim$100~mas, that is, close to the bright blobs detected with PCA analysis. 
Both rings are inhomogeneous; in particular, the one at 180~mas shows a decrease in the brighness around PA=$45^\circ$ compared to the surrounding ring signal (at $20^\circ$ and $80^\circ$), and there are several brightness enhancements in the North-West and South-West directions. The ring appears more clearly in the April 2016 image, possibly due to a better quality of the data and a larger rotation field. The inner ring at 100~mas is quite bright with a brighter region in the North West direction in the 2015 image. However, it is not detected in each reduction (in particular, it is hardly seen in RDI data without coronagraph), and its appearance depends on the scale used (see Sec.~\ref{sec:Structure}). It appears much less bright in the 2016 April image, although we still detect a signal.

\section{An inhomogeneous ring at 180 mas}
\label{sec:Blobs}

\subsection{Simulation of cADI reduction with PDI data}

\begin{figure*}
	\includegraphics[width=0.9\textwidth]{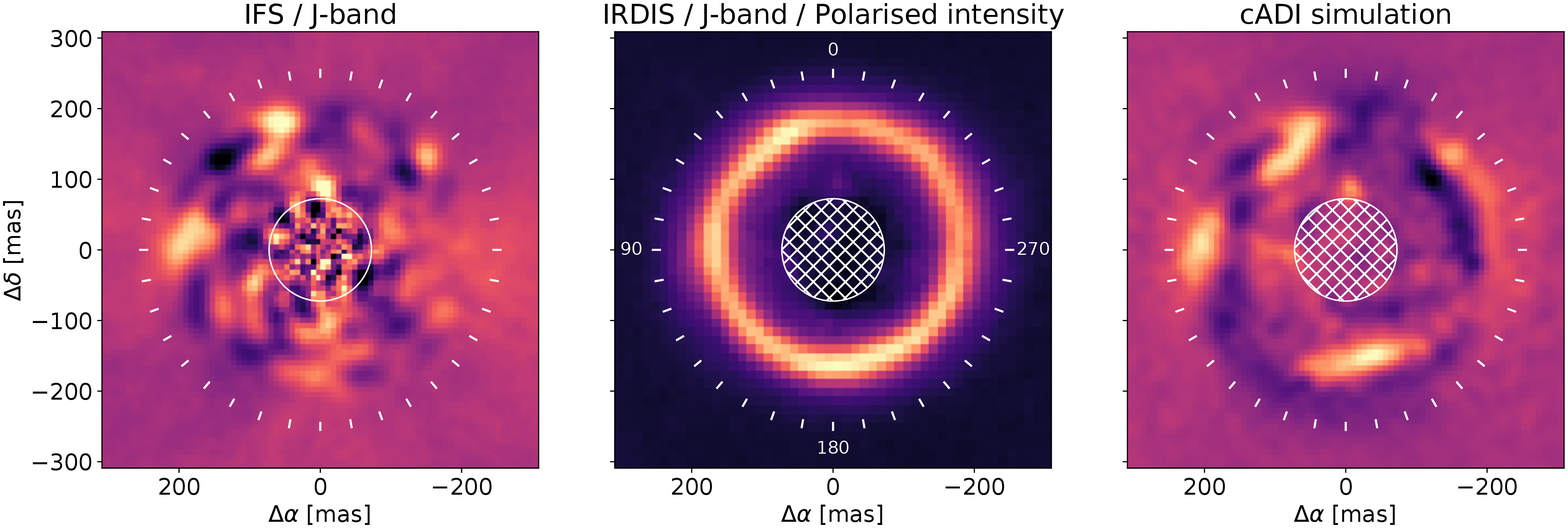}
	\caption{\textit{Left}: IFS image in $J$ band with 50 PCA modes subtracted. \textit{Middle}\,: IRDIS PDI polarized intensity image in $J$ band, which shows a bright irregular ring. \textit{Right}: Result of a cADI simulation using the polarized intensity image as input (see text for details). The circular grid at the centre represents the centre star covered by coronagraphic mask in the PDI data. The corresponding position in the IFS data is also represented in the IFS $J$ band image, although this data was obtained without a coronagraph (empty circle). Tick marks placed every 10$^{\mathrm{o}}$ in position angle are plotted outside of the area of interest. North is up and east is left.}
	\label{fig:simu}
\end{figure*}

\begin{figure}
\centering
	\includegraphics[width=0.40\textwidth]{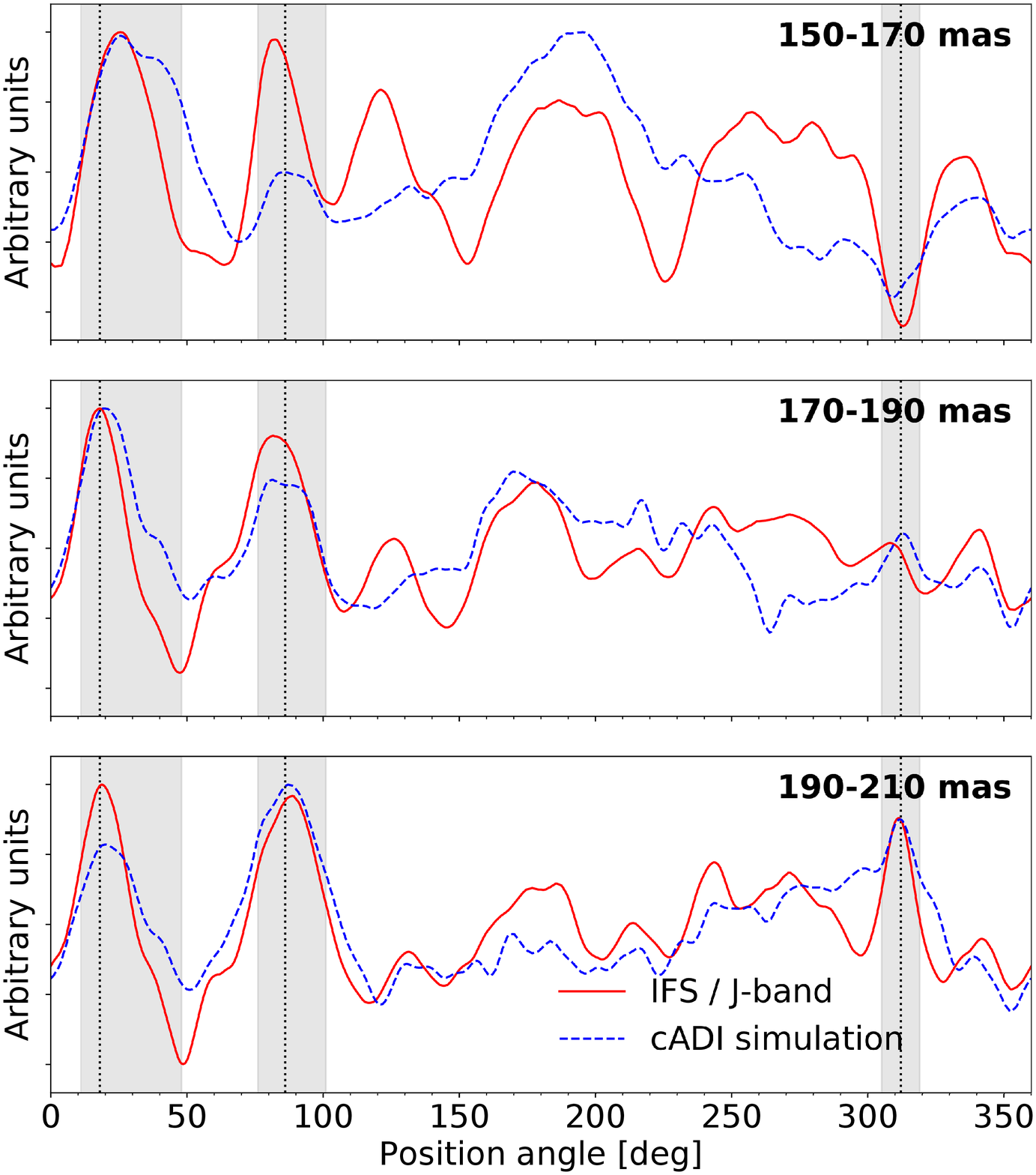}
	\caption{Azimuthal cuts at three different separations in the IFS $J$ band data with 50 PCA modes subtracted (plain red line) and in the cADI simulation (dashed blue line). The separation is indicated in the top right corner of each plot. The signal is measured every 0.5$^{\mathrm{o}}$ and averaged in slices of 20~mas along the radial dimension. The curves are normalized to arbitrary units allowing comparison between them. Grey areas highligth the structures that show the strongest correlation, with the dashed lines indicating the position of highest signal for a given separation.}
	\label{fig:cadi_simulation_profile}
\end{figure}

An important indication for understanding the nature of the detected structures would be to know if the scattered light is polarized. Indeed, planets are usually considered to not emit polarized light, contrary to protoplanetary disks \citep[see however][who suggest that very small polarization is possible in some cases]{Stolker2017}. Polarized light due to reflection from hot Jupiter planets could also be detected in the optical (UBV bands). However, the produced signal would be low \citep{Berdyugina2008, Berdyugina2011} which might be impossible to detect when the planet is embedded in a disk, which produces polarized scattered light.

To investigate the nature of the blobs at $\sim$180~mas, we use IRDIS PDI (Polarimetric Differential Imaging) data that were acquired on 2015-05-02 with the \texttt{ALC$\_$YJ$\_$S} apodized-pupil Lyot coronagraph (145~mas in diameter) in the $J$ band and reduced following \cite{deBoer2016}. A full analysis and modelling of the PDI data will be presented in a forthcoming paper \citep{Pohl2017b}. The left and middle panels of Fig.~\ref{fig:simu} present a comparison between the IFS $J$ band data and the IRDIS PDI data of the very central region around the star ($\pm$300~mas). The ring at 180~mas in the IRDIS polarized intensity image is detected at extremely high significance and the use of a small coronagraphic mask allows to unambiguously confirm the existence of the cavity inside of the ring. 
The polarized intensity image also clearly shows a variation of the ring brightness as a function of the position angle, with an increase of the brightness at positions angle of $\sim$20$^{\mathrm{o}}$, $\sim$90$^{\mathrm{o}}$, $\sim$180$^{\mathrm{o}}$ and to a lesser extent at $\sim$310$^{\mathrm{o}}$. The brightness of the structure has been measured by \citet[][Fig.~3]{Pohl2017b}, and higher signals at these same PA are clearly visible. Interestingly, these regions of increased brightness seem to correspond to position angles where the IFS $J$ band image shows extended bright structures at a significance of 2.5--3$\sigma$ above the surrounding residuals\footnote{Note that the signal-to-noise maps were calculated on our images assuming that we were looking for point sources, which does not translate directly when considering extended structures such as disks. However, the structures visible in the data are clearly identifiable above the surrounding background.}. In particular, the bright structures at PA$\approx20^\circ$ and $310^\circ$ in PDI data seem to correspond to the structures shown by the blue arrows in Fig.~\ref{fig:detectionIRDIFS} (structures A and B), and a structure at PA=380$^{\circ}$ and $\rho$=100 mas seems to correspond to structure C. A bright extended structure also appears between 180$^{\circ}$ and 210$^{\circ}$ in the cADI simulation images, which is clearly visible in the PDI image but appears fainter in the IFS $J$ band image. Finally, we notice the strong similarity between the ring appearing in the PDI data and in the RDI data from 2016 April (Fig.~\ref{fig:RDI}). \\

To confirm that the residual structures seen in the IRDIS and IFS data are in fact ring structures filtered by the ADI processing, we perform a simulation of ADI reduction using the IRDIS PDI polarized intensity image. First, we create a data cube with 1709 copies of the PDI image (because this is the number of frames after selection using the LAM-ADI pipeline, see Sec.~\ref{sec:PCAreduction}), corresponding to each of the IFS images, with each of the images being rotated to match the pupil offset rotation and the position angle of the observations. Then, the median image of the cube along the temporal dimension is calculated and subtracted to each of the images in the cube (classical ADI). Finally, all the images are rotated back to a common orientation and mean-combined. The result of this simulation is presented in the right panel of Fig.~\ref{fig:simu}. \\
Visually, we see a strong correlation between the main structures identified in the IFS $J$ band image and the bright ring in the disk that have been spatially filtered by the ADI analysis. The effect of ADI processing on disks has already been studied by \citet{Milli2012} and they have identified that ADI can have a strong impact on the flux and morphology of disks, up to the point of creating artificial features. This effect has also been encountered in the case of HD\,100546 \citep{Garufi2016} and T\,Cha \citep{Pohl2017a}. The ring of HD\,169142 seen almost face-on is an extreme case: all azimuthally symmetric structures of the ring are completely filtered out by ADI, leaving only the signature of the features brighter or fainter than average. In the simulation, the shape of the features at $\sim$20$^{\mathrm{o}}$ and $\sim$90$^{\mathrm{o}}$ is almost identical to that in the IFS image. The same bright spot at a position angle of $\sim$310$^{\mathrm{o}}$ is also clearly visible. 

For a more quantitative assessment, we compare in Fig.~\ref{fig:cadi_simulation_profile} azimuthal cuts of the IFS $J$ band data with 50 PCA modes and the cADI reduction simulation, measured at different separations from the star. The signal is averaged in annuli of 20~mas of radial extension to smooth the small pixel-to-pixel variations in the data. These cuts show a very strong correlation between some of the main features seen in the IFS data and the cADI reduction simulation: the Pearson correlation coefficients between the two data sets are equal to 0.47, 0.64 and 0.80 for the 150--170~mas, 170--190~mas and 190--210~mas azimuthal cuts respectively. This correlation is calculated on the full ring, but it would be even higher if we considered more local correlations centered on the main features. \\

\subsection{Interpretation of the results}

\begin{figure*}
$\begin{array}{cc}
	\includegraphics[width=0.45\textwidth]{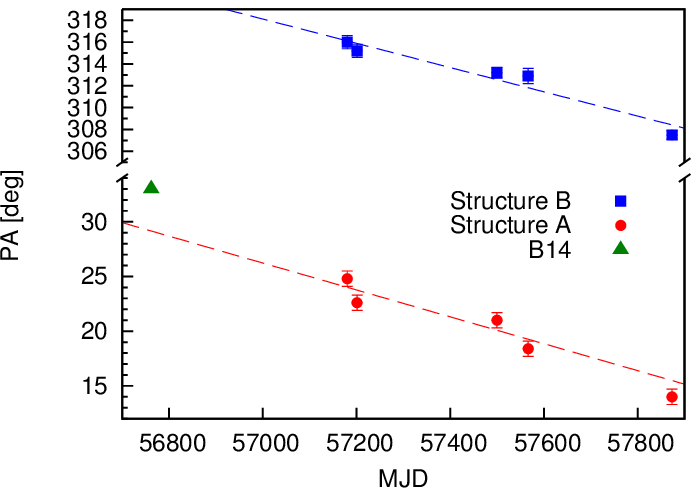} &
	\includegraphics[width=0.43\textwidth]{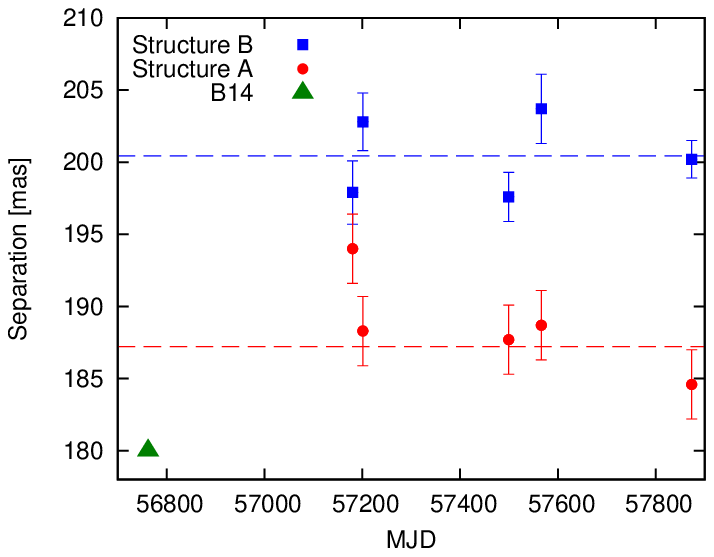}
\end{array}$
	\caption{\textit{Left}: PAs of the structures A (red circles) and B (blue squares) as a function of MJD. \textit{Right}: Separations of the blobs. The dashed lines show the average separation and the corresponding Keplerian speed. The green triangle is the position given by \citet{Biller2014}.}
	\label{fig:positionsAB}
\end{figure*}

In the NIR, we are not only sensitive to the thermal emission from point sources but also to stellar light scattering of the protoplanetary dust. The detected point source at 180~mas \citep{Biller2014} lies on the surface of the ring which is optically thick in the NIR (as the whole disk). Thus, the signal of a point source in the midplane of the disk will be dominated by scattered light and the planet emission would be strongly absorbed by the dust, making impossible its detection. It would be possible to detect a very massive companion at that location, but in that case we would expect it to have opened a deep and broad gap. Instead, we find a ring. Since the detection of polarized light from a planet is not expected, we conclude that the blobs that we detect both in PDI and non-polarized data are part of the same structure: the ring. Hence, we can conclude with a high confidence level that our images show disk features rather than planetary companions for the structures A and B. We can thus exclude the thermal emission from giant planets being consistent with the blobs signal, but we cannot exclude clumps in an early stage of planet formation.

Multiple blobs are found on the same orbit at $\rho\approx180$~mas. Of particular interest is the structure at PA$\approx20^\circ$ (structure A in Fig.~\ref{fig:detectionIRDIFS}) because it is bright and appears both in PDI and non-polarized data. We try to investigate if it corresponds to an object candidate, and in particular, to the one detected by \cite{Biller2014} at PA=33$^\circ$ and $\rho$=180~mas (see Sec.~\ref{sec:Introduction}) because it is close to our current detection's position.
Considering the stellar distance ($117.3^{+3.8}_{-4.1}$~pc) and mass (1.65~M$_\odot$) of HD\,169142 (see Tab.~\ref{tab:param}), and the inclination of the disk \citep[13$\pm1^{\circ}$,][]{Panic2008}, \cite{Biller2014}'s candidate should have an orbit of $\sim$78.5~yr if in the disk plane with no eccentricity, and therefore should have moved of $\sim$13.7$^{\circ}$ from June 2013 to June 2016 . This would bring it to a PA=19.3$^{\circ}$ if in a clockwise motion, which is in very good agreement with the measured PA of our structure (20$^{\circ}$).

The movement of the blobs (structures A and B) according to the different epochs of observations are shown in Fig.~\ref{fig:positionsAB}. We also include the separation and PA obtained by \cite{Biller2014} (Fig.~\ref{fig:positionsAB}, green triangle). For each structure, we calculate the average separation over all the epochs \citep[including the position of][for the calculation of the separation of structure A]{Biller2014}, and the Keplerian speed corresponding to this average separation. We then plot the Keplerian speed on the PA figure. We can see that the structure motions are compatible with Keplerian speeds. We note that \cite{Biller2014}'s position is indicated without error bars, but they give a rough estimation of the position in their paper.
We already showed that our detections could be related to blobs in the disk. We thus conclude that the PA and separation evolution of the blob are consistent with a Keplerian motion. The blobs trace the bright ring in the disk, and rotates in the clockwise direction with a Keplerian velocity. Moreover, ALMA data \citep{Fedele2017} provide the direction of rotation of the disk (the Northern part of the disk is moving faster toward us than the local rest frame, while the Southern part is moving slower) and its closest side to the observer (the Western side), which are compatible with the blob clockwise motion. 

The exact nature of the blobs and the origin of the bright disk rings and gaps in general remain to be investigated. We only make a hypothesis here which scenarii could be compatible with our observations. We also refer the reader to the upcoming work by \cite{Pohl2017b} for a detailed modeling study of the disk around HD169142 including planet-disk interaction processes and dust evolution dynamics.\\
The first possibility invokes intrinsic disk variations in density and temperature. Indeed, the dust concentration in the ring might be a tracer of a maximum density in the gas profile. This jump in the surface density could trigger the formation of vortices by the Rossby Wave Instability (RWI) which concentrates dust azimuthally. Our observations look like figs. 3 and 5 of \citet{Meheut2012} who display simulations of Rossby vortices with several irregular blobs of enhanced dust density on the same orbit. \citet{Barge1995} show that these vortices could be favourable places to initiate planets formation. If this is the case, HD169142 could be the site of ongoing planet formation, at an earlier stage than previously expected.
Although multiple vortices are non-permanent states, the mass ratio between the disk and star \citep[$0.03$, considering the refined estimate of the disk mass by][]{Monnier2017} does not make the disk gravitationaly unstable, and woud allow self-gravity to improve the stability of multiple vortices \citep{Lin2011}. 
If the blobs in our image indeed trace vortices, we would only observe their signatures in the upper disk layers in our SPHERE observations. 
While ALMA observations, that trace the midplane layers, have already been interpreted as vortices in protoplanetary disk \citep[see e.g.][]{vanderMarel2013}, recent observations of HD\,169142 with this instrument \citep{Fedele2017} did not show any asymmetrical features at a resolution of 0.2-0.3". 
In addition, the spatial distribution of particles inside a vortex depends on their size \citep[see e.g. ][]{Lyra2009, Gonzales2012} and such structures would thus appear differently in the sub-millimeter regime. However, the currently available ALMA observations would not be able to resolve the various structures shown in this paper if they have the same or smaller spatial extent.

The second scenario involves illumination variations because of azimuthally asymmetric optical depth variations through an inner disk closer to the star. Even if much less plausible, this scenario has already been raised in previous studies concerning HD169142 \citep{Pohl2017b,Quanz2013}. The observed brightness variations at 180~mas are relatively small [\cite{Pohl2017b} suggest an azimuthal brightness variation of 25$\%$ in the PDI data] and could be caused by such variations. Besides, the inner disk at $\sim0.3$~au is known to present a variable spectral energy distribution (SED) in the NIR. \cite{Wagner2015} propose several scenarios to explain the variations of the SED of HD169142, but they invoke stable shadowing effect, otherwise an anti-correlated variability in the emission of the inner and outer disks should be observed in the SED, which is not the case.
If additional material exists within our inner working angle, at high altitude, it could shadow the ring at 180~mas, but this remains to be investigated. In any case, our discovery of the Keplerian movement of the structures at 180~mas strongly suggests that the origin of their intensity variation does not come from an inner structure.

\section{A point-like structure at 100 mas}
\label{sec:Structure}

In the data we also detect a point-like structure North of (PA$\approx4^\circ$) and close to ($\rho$=105$\pm$6~mas) the star (structure C in Fig.~\ref{fig:detectionIRDIFS}). This position was determined using the ASDI-PCA algorithm \citep[][see Sec.~\ref{sec:PCAreduction}]{Mesa2015}. This structure is persistent in both IRDIS and IFS data, and is visible with large range of PCA reductions subtracting 12 to 200 modes. The separation of this structure from the star is similar to the separation of the object detected by \cite{Biller2014} with NaCo in $L'$ band, and is slightly offset, but consistent with \cite{Reggiani2014}'s detection.

To confirm the robustness of our detection, we first split the IRDIS and IFS data into two sub-sets using the LAM-ADI pipeline, which were analysed following the same procedure as described in Sec.~\ref{sec:PCAreduction}. In each of the resulting images the structure was still visible at a S/N higher than the surrounding background. This reduction shows a structure at PA=355$\pm3^{\circ}$ and $\rho=93\pm6$~mas in average over all wavelengths (S/N of 3.3 in $H$ band), which is consistent with the estimation provided with the ASDI-PCA pipeline. 
The IRDIS data were also analysed with the \texttt{PYNPOINT} pipeline \citep{Amara2012}. In this analysis the structure is marginally detected, as it only appears between 8 and 15 PCA modes (over 50). 

The structure at $\sim$100~mas appears in most data set as a somewhat extended structure (see Fig.~\ref{fig:detectionIRDIFS}), in particular in the $H$ and $K$ bands. This was not detected previously in \cite{Biller2014} and \cite{Reggiani2014} analysis, where it appeared point-like at $L'$ band and was not detected in lower sensitivity, short wavelength observations . 

The structure is partially visible in the PDI image, and in the simulated image of cADI reduction of PDI data (Fig.~\ref{fig:simu}). This means that its signal is polarized, which indicates light scattered by dust than the emission from a planet photosphere. The position of this structure in the simulation of cADI reduction is $\rho$=82$\pm$3~mas at PA=355$\pm2^{\circ}$, which is very similar to the PA estimate from the LAM-ADI pipeline. The separation measured in the simulation remains within the error bars of the estimate position in the IFS $J$ band ($\rho$=90.5$\pm$2.5~mas, PA=357.9$\pm3.0^{\circ}$), but the separation is smaller than the estimate obtained with the ASDI-PCA algorithm. This could be explained by the measured positions on the real images which are made in average over all wavelenghts, as for the LAM-ADI pipeline.

As seen in Sec.~\ref{sec:Blobs}, the average brightness of a disk can be filtered out by the cADI reduction. It is thus possible that this structure coming out of the simulation of cADI reduction of PDI data actually traces a yet undiscovered ring, and that this structure is a bright part of this ring. Moreover, the structure lies close to the edge of the mask, so it is likely attenuated in the PDI image. This may explain why this hypothetical ring is not detected in the PDI image.
The RDI image also shows a ring at $\sim$100~mas, that is, very close to the coronagraph (Fig.~\ref{fig:RDI}). Moreover, the ring is not retrieved in each detection, appearing sometimes in mean-scaled images, other times in median-scale images.  
These results tend toward a marginal detection of a bright ring at a separation of $\sim$100~mas, that our current observations unfortunately cannot confirm. Additional PDI observations closer to the star without coronagraph would certainly bring precious information.

\section{Summary and Conclusion}
\label{sec:conclusion}

We performed observations of the Herbig Ae star HD\,169142 using SPHERE/VLT in the NIR domain with and without coronagraph to investigate the inner parts of the system (<300~mas). We observed this star at five different epochs, leading to several new results\,:
\begin{itemize}
\item The ADI analysis show bright structures both in IRDIS ans IFS data. These structures appear more extended in the $H$ band than in the $YJ$ and $K$ bands. They are mainly located at separations of $\sim$180-200~mas and $\sim$100~mas.
\item The RDI reduction clearly shows a bright ring at 180-200~mas. It also shows a hint of another inner ring located at $\sim$100~mas. However, it is very close to the edge of the coronagraph, and does not appear identical in every data treatment. Thus, it cannot be confirmed.
\item To assess the origin of the structures seen in ADI reductions, we performed a cADI simulation using the image of the ring detected in PDI at 180 mas. While the main component of the ring is filtered out we still observed residual structures that appear to be common to both PDI and ADI reductions. We therefore conclude that these structures are actually bright parts of the disk. 
\item Given that (i) the bright blobs seen in PCA analysis, in particular structures A and B, and the ring detected with RDI analysis are located at the same separation (180~mas), and (ii) these blobs and the polarized data are actually part of the same structure: the ring, we conclude that the bright blobs trace this bright ring in the disk.
\item From the previous result, and considering the stellar parameters, we demonstrate that the structure A follows a Keplerian motion along the ring. Considering this movement, structure A is very likely to be the same structure as the one detected by \cite{Biller2014} at PA=33$^{\circ}$ and $\rho=180$~mas. It is likely that \cite{Biller2014} actually detected a bright structure in this ring, and that the ring brightness was averaged following the same process as for our PCA treatment. The structure B also shows a Keplerian movement and also traces the bright ring in the disk. The latter thus rotates in a clockwise direction with a Keplerian velocity, with the Western side closer to us and the Eastern farther.
\item The ring at 180~mas shows an inhomogeneous brightness. One explanation could involve Rossby vortices before they merge into one bigger vortex. These vortices are ideal place to trigger planetary formation at an early stage. If the inner ring is real, another explanation could be illumination effects from this inner ring. The irregularity of this ring could produce azimuthally optical depth variations of the ring at 180~mas, but the angular velocity does not match this hypothesis.
\item The structure located at 100~mas (structure C) appears to be point-like at shorter and longer wavelengths but extended in the $H$ band, and its position is consistent with previous $L'$ band detections. The RDI images show a possible inner ring at the same separation. Thus, although marginally detected, it could also trace a yet undetected ring that is even closer to the star. The PCA treatment could easily make it appear point-like, as it does for structures A and B.
\end{itemize}
HD\,169142 is a very interesting case to study planet formation as it is a pre-transitional disk showing a succession of bright rings, gaps and a a ring/gap alternation. To confirm the inner ring, additional observations would be needed, but the resolution of actual (and even future) direct-imaging instruments would hardly allow such a discovery.

\section*{Acknowledgements}

R. L. thanks CNES for financial support through its post-doctoral programme. This work has been carried out in part within the frame of the National Competence in Research (NCIR) "PlanetS", supported by the Swiss National Science Fundation (SNSF). The authors thank the ESO Paranal Staff for support for conducting the observations. We also warmly thank H. M\'eheut, for the useful discussions about vortices. We acknowledge financial support from the Programme National de Plan\'etologie (PNP) and the Programme National de Physique Stellaire (PNPS) of CNRS-INSU. This work has also been supported by a grant from the French Labex OSUG@2020 (Investissements d'avenir - ANR10 LABX56). The project is supported by CNRS, by the Agence Nationale de la Recherche (ANR-14-CE33-0018). MB acknowledges funding from ANR of France under contract number ANR-16-CE31-0013 (Planet Forming Disks). This work has made use of the SPHERE Data Centre, jointly operated by OSUG/IPAG (Grenoble), PYTHEAS/LAM/CeSAM (Marseille), OCA/Lagrange (Nice) and Observatoire de Paris/LESIA (Paris). We thank P. Delorme and E. Lagadec (SPHERE Data Centre) for their efficient help during the data reduction process. SPHERE is an instrument designed and built by a consortium consisting of IPAG (Grenoble, France), MPIA (Heidelberg, Germany), LAM (Marseille, France), LESIA (Paris, France), Laboratoire Lagrange (Nice, France), INAF-Osservatorio di Padova (Italy), Observatoire astronomique de l'Universit\'e de Gen\`eve (Switzerland), ETH Zurich (Switzerland), NOVA (Netherlands), ONERA (France) and ASTRON (Netherlands) in collaboration with ESO. SPHERE was funded by ESO, with additional contributions from CNRS (France), MPIA (Germany), INAF (Italy), FINES (Switzerland) and NOVA (Netherlands). SPHERE also received funding from the European Commission Sixth and Seventh Framework Programmes as part of the Optical Infrared Coordination Network for Astronomy (OPTICON) under grant number RII3-Ct-2004-001566 for FP6 (2004-2008), grant number 226604 for FP7 (2009-2012) and grant number 312430 for FP7 (2013-2016).

We acknowledge the use of the electronic database from CDS, Strasbourg and electronic bibliography maintained by the NASA/ADS system. 

This work has made use of data from the European Space Agency (ESA) mission {\it Gaia} (\url{https://www.cosmos.esa.int/gaia}), processed by the {\it Gaia} Data Processing and Analysis Consortium (DPAC, \url{https://www.cosmos.esa.int/web/gaia/dpac/consortium}). Funding for the DPAC has been provided by national institutions, in particular
the institutions participating in the {\it Gaia} Multilateral Agreement.

\bibliographystyle{mnras}
\bibliography{Sphere}

\appendix

\section{Signal-to-noise maps}
\label{sec:AppendixA}

\begin{figure*}[h]
	\centering	
	\includegraphics[width=0.8\textwidth]{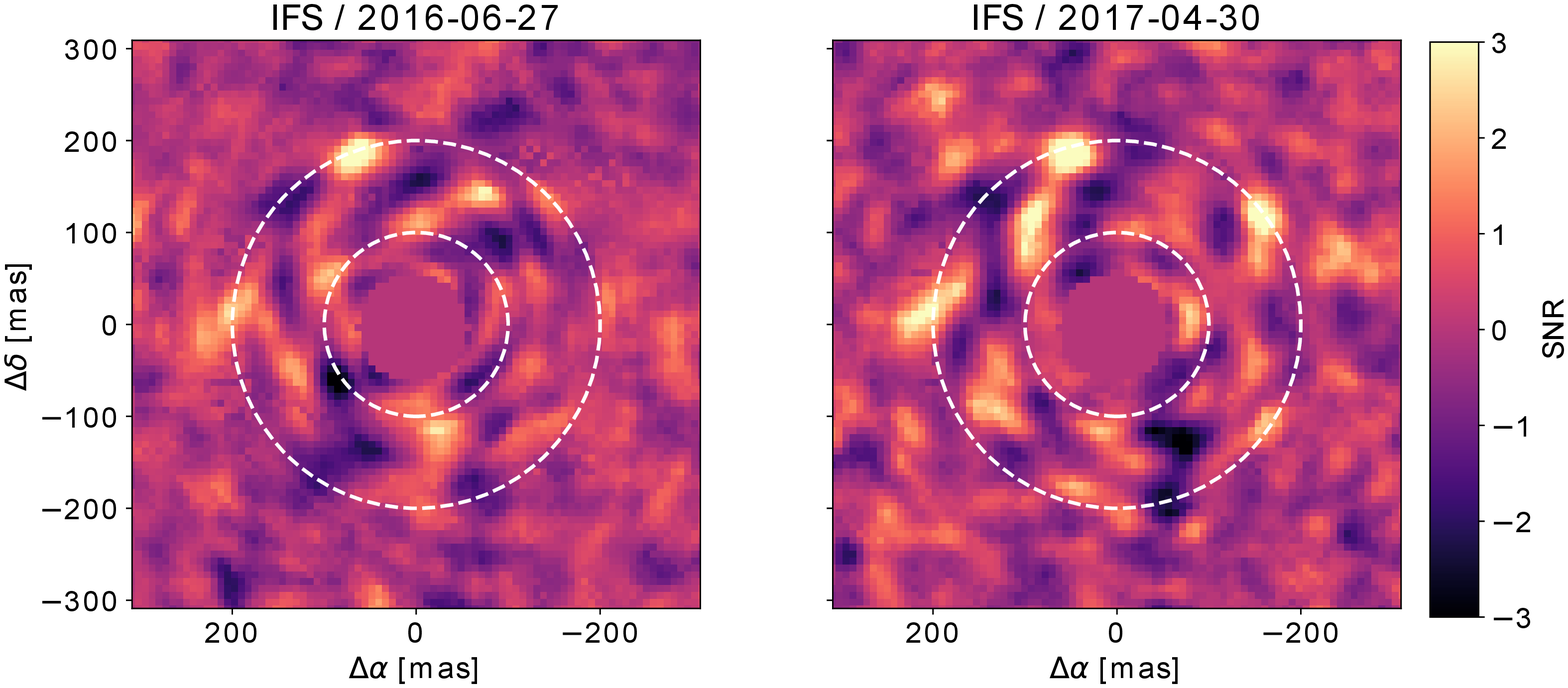}
	\caption{Maps of the S/N of the IFS data from 2016 June 27 (left) and 2017 April 30 (right). The S/N was calculated following the method described in \citet{Zurlo2014}. The rings have radii of 100~mas and 200~mas. The highest signals appear at the same locations as the ones in Fig.~\ref{fig:detectionIRDIFS}. North is up and east is left.}
	\label{fig:snrmap}
\end{figure*}


\label{lastpage}
\end{document}